# Primary activity measurement of an Am-241 solution using microgram inkjet gravimetry and decay energy spectrometry

Ryan P. Fitzgerald, Bradley Alpert, Denis E. Bergeron, Max Carlson, Richard Essex, Sean Jollota, Kelsey Morgan, Shin Muramoto, Svetlana Nour, Galen O'Neil, Daniel R. Schmidt, Gordon Shaw, Daniel Swetz, and R. Michael Verkouteren

*National Institute of Standards and Technology, 100 Bureau Drive, Gaithersburg, MD USA*

Abstract

We demonstrate a method for radionuclide assay that is spectroscopic with 100 % counting efficiency for alpha decay. Advancing both cryogenic decay energy spectrometry (DES) and drop-on-demand inkjet metrology, a solution of Am-241 was assayed for massic activity (of order 100 kBq/g) with a relative combined standard uncertainty less than 1 %. We implement live-timed counting, spectroscopic analysis, validation by liquid scintillation (LS) counting, and confirmation of quantitative solution transfer. Experimental DES spectra are well modeled with a Monte Carlo simulation. The model was further used to simulate Pu-238 and Pu-240 impurities, calculate detection limits, and demonstrate the potential for tracer-free multi-nuclide analysis, which will be valuable for new cancer therapeutics based on decay chains, Standard Reference Materials (SRMs) containing impurities, and more widely in nuclear energy, environmental monitoring, security, and forensics.

## 1. Introduction

We present a new spectroscopic capability for quantifying radionuclides in solution. Radionuclide activity standards must be based on primary reference measurements, meaning that the methods are internally consistent and not dependent on any measurement standard for a quantity (i.e., activity) of the same kind [1]. Existing primary methods for radioactivity measurements are generally not spectroscopic, so rely on secondary measurements by other methods to identify, quantify, and place limits on radionuclidic impurities that can cause bias in the primary measurement [2]. For example, the $^{229}$Th massic activity standardization referenced in [2] had a combined standard uncertainty of only 0.30 % but required a 1.5 % correction for $^{228}$Th impurity [3]. We demonstrate a new primary method for measuring activity of microgram quantities of aqueous samples that may contain multiple radionuclides, such as environmental samples, cancer therapeutics based on decay chains, and samples containing radionuclidic impurities. Here, we perform a primary standardization of Am-241 massic activity and place limits on alpha-emitting impurities.

The new approach is based on quantum devices called transition-edge sensors (TES). The TES is operated near the transition temperature $T_c$ between the superconducting and normal states of a superconducting film, where the resistance of the sensor is highly sensitive to temperature [4]. The principles of TES operation are illustrated in the figures in [5]. In recent years, TESs have become familiar in multiple fields of science from space-based astronomy to laboratory atomic, ion, and plasma physics where they have radically improved sensitivity and resolution in x-ray spectroscopy [5]. Decay Energy Spectrometry (DES) using TES and other cryogenic sensors (such as magnetic microcalorimeters [MMCs]) has great potential for absolute activity measurements for radionuclides decaying by alpha [6], beta [7], electron capture [8], and mixed [9] decay modes, achieving spectral resolving powers of up to 5000 [10]. In DES, the energy released by a single radioactive decay is captured in a $4\pi$ absorber (made of Au for the present work) and measured with a sensitive thermometer. The total energy of the nuclear decay is referred to as the $Q$-value, so that DES is also referred to as $Q$-spec [10]. Among other efforts, a European consortium called Prima-LTD has achieved significant progress toward measuring absolute activity with low-temperature detectors [11].

Here, we report advances in DES measurements, pulse processing with live-timed spectroscopic analysis, and drop-on-demand inkjet gravimetry that enable a fully-realized DES-based standardization of an Am-241 solution for massic activity (of order 100 kBq/g), achieving uncertainties that are competitive with established primary methods.

## 2. Method

### 2.1 Overview

The overall vision for measuring massic activity by DES was described in a previous paper [12] . An ensuing paper described the first experimental results for a preliminary, hand-operated micro-drop dispensing method [13].

Figure 1 of this work illustrates the source preparation scheme in which a stock solution of $^{241}$Am was gravimetrically dispensed by pycnometer into both an LS source and a small vial that served as a reservoir for the inkjet. The inkjet device draws solution from the reservoir and dispenses it into prepared LS vials and onto nanoporous-gold-coated Au foils. The foils are sitting on gel "placemats" that are later measured by LS to ensure no loss of activity during dispensing [13]. After the inkjet depositing is complete, the reservoir is removed from the inkjet device and re-sampled by a pycnometer into another prepared LS source to verify consistency of the activity concentration during the deposition process.

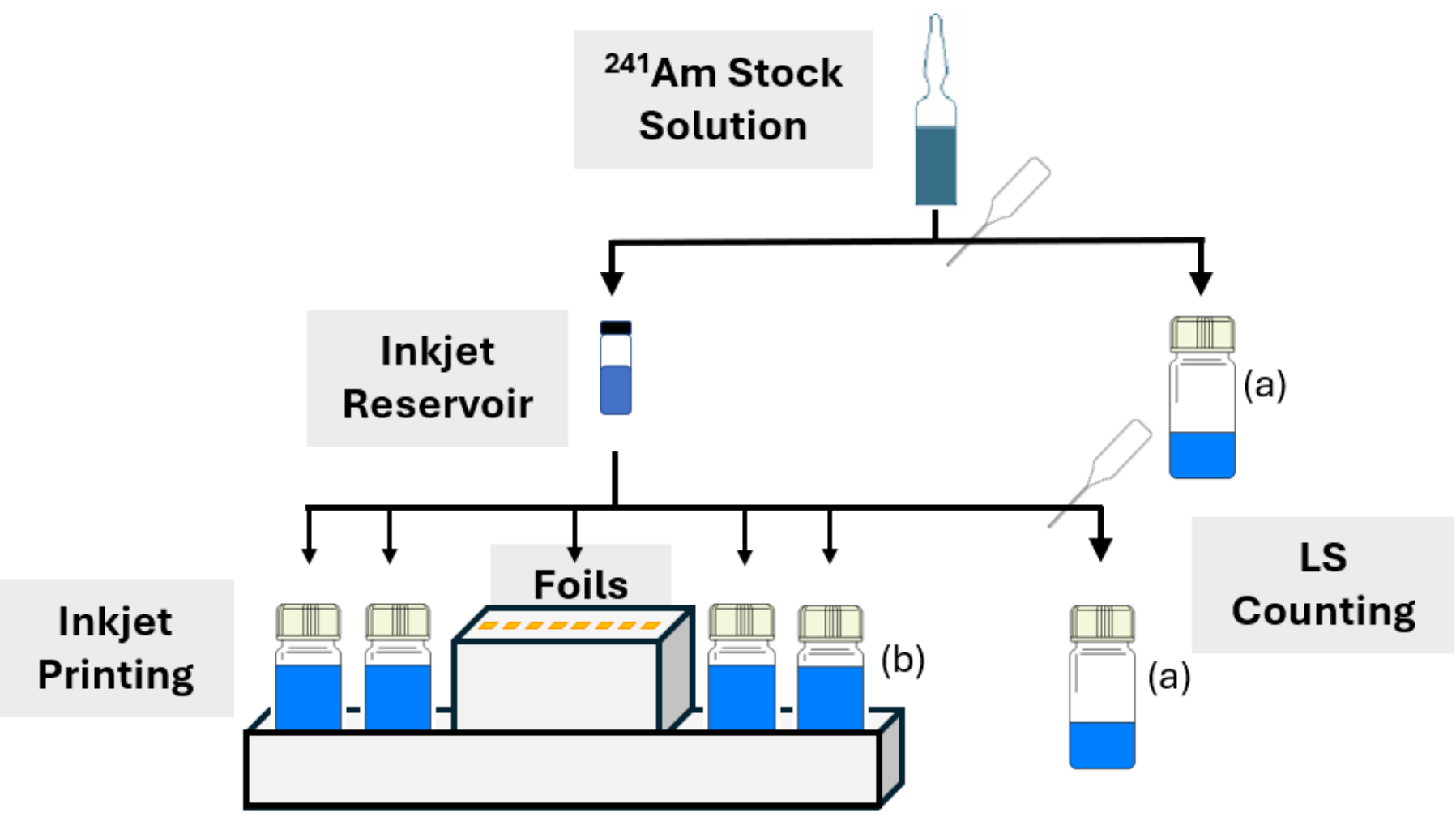

Figure 1 Overview of gravimetric source preparation of LS sources by both pycnometer and inkjet and DES sources (Au foils) by inkjet. The foils are mounted on gel placemats during dispensing.

After the Au foils dry, they are folded in half and rolled flat using a manual roller press to encapsulate the radioactive material, then mounted on a TES chip. Figure 2 shows the geometry of the TES chip and a photograph of a chip with a source attached. The chip is designed with a central island that is thermally linked to the cryostat by meandering silicon “legs”. The thermal conductivity of this link contributes to pulse shape and energy resolution. On the central island, the TES, made from a Mo-Cu bilayer, is thermally connected to the source through a gold pad onto which the source is placed. The connection is improved using a thin layer of Ga-In eutectic. The island is heated to a temperature near $T_c$ by a bias current that is conducted through Mo traces, which also provides readout through a two-stage SQUID (superconducting quantum interference device) amplifier.

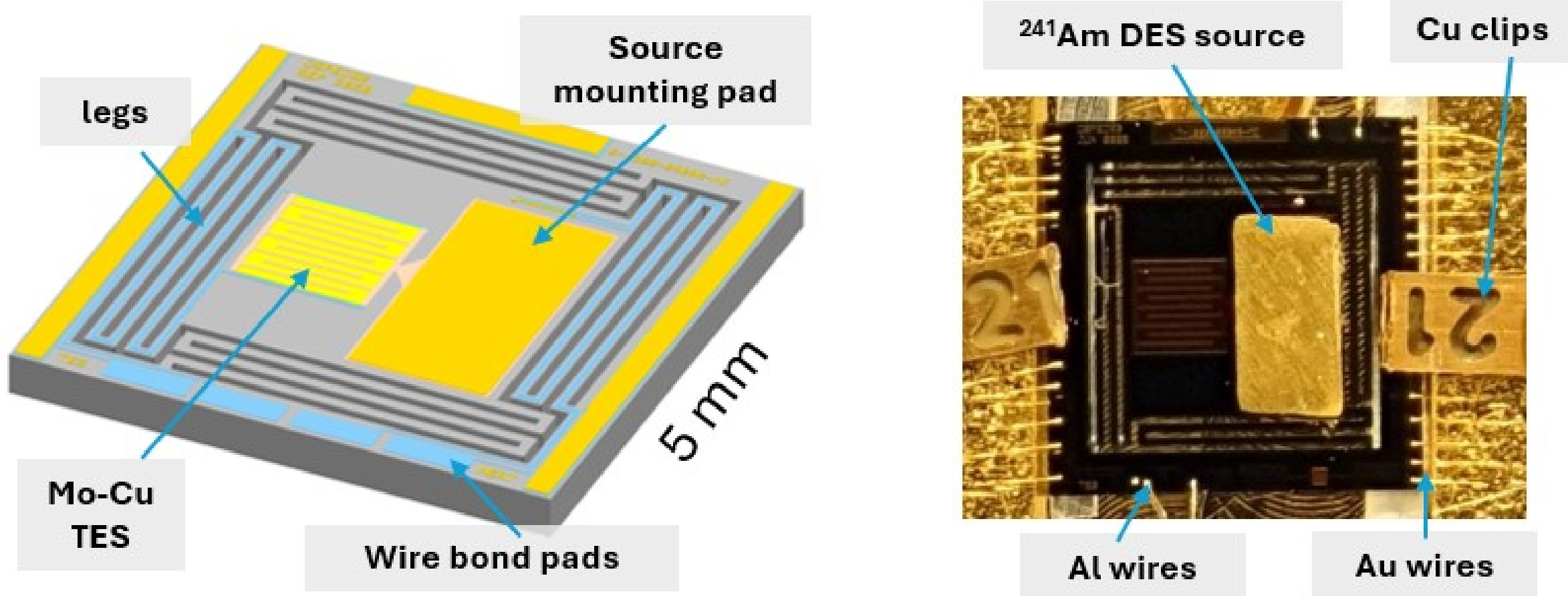

Figure 2 Design of the TES chip (left) fabricated on a silicon wafer and consisting of a Mo-Cu TES, silicon legs (variable number of legs depending on choice of thermal conductivity), Mo wire traces and pads for wire bonding to the TES bias and SQUID readout, and a gold pad in thermal contact with the TES for mounting the DES source. A photo (right) of a TES chip with a folded and rolled source mounted on the pad using a thin layer of In-Ga eutectic. Copper clips and gold wires mount the chip to a gold-coated copper box that is held at the 50 mK base temperature of the dilution fridge. Al wires make the connection to TES bias and SQUID readout. See section 2.3.

The heat absorbed by the source from a decay event is dissipated through the legs and through electro-thermal feedback [4]. The feedback signal of the SQUID array is read at room temperature by an ADC. The resulting pulse, Figure 3, is analyzed for pulse height, as described in section 2.4. A histogram of pulse heights is used to generate a decay energy spectrum, which is analyzed for count rate in particular regions of interest. Monte Carlo simulations assist in assigning radionuclides to spectral features and decomposing spectra that overlap in the case of multiple radionuclides.

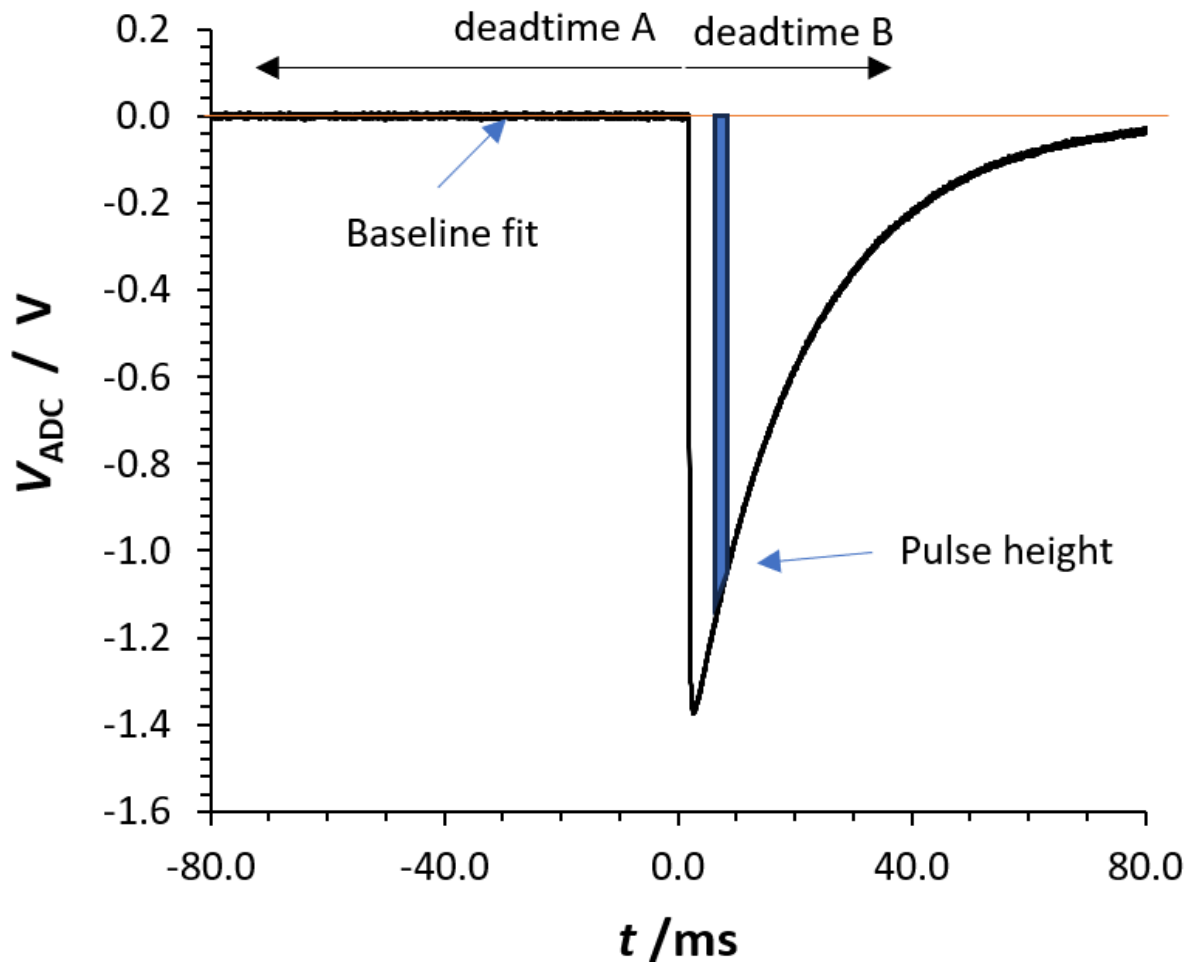

Figure 3 Example DES pulse. The voltage ($V_{ADC}$) measured at the ADC shows a negative pulse following a decay event. The baseline voltage is fit prior to a pulse and extrapolated under the pulse. The pulse height is measured at a fixed time after the pulse starts. Deadtimes before and after a pulse are applied to reduce pileup. See Section 2.4 for more details.

## 2.2 Materials

The stock solution of Am-241 was in ≈ 1 mol/L nitric acid having nominal massic activity of $1.5 \cdot 10^5$ Bq/g at the reference date 1 September 2022. Extremely small aliquots of solution (< 10 µg) were dispensed (Section 2.2) for cryogenic DES measurements (Section 2.3). Larger (≈ 50 mg) aliquots were dispensed by traditional methods for validation measurements using non-spectroscopic liquid scintillation counting (Section 2.4).

For this first demonstration, Am-241 was chosen as it decays purely by alpha decay (rather than beta decay or electron capture), mostly to the 59.5 keV excited state of Np-237 [14, 15]. The resulting alpha particle and Np-237 recoil nucleus come to rest in the gold absorber, depositing their kinetic energies. The Np-237 deexcitation results in emission of conversion electrons, Auger electrons, X-rays, and $\gamma$-rays. The 59.5 keV $\gamma$-ray is emitted with a probability of 35.9 % and has a significant probability of escaping the absorber foil without depositing energy. This escape provides a useful test of our spectral analysis methods, as it can interfere with Pu-238 events in the energy spectrum.

The Am-241 solution was deposited (Section 2.2) onto a layer of nanoporous gold (npAu) film that had been deposited on 15 µm or 25 µm gold foils to create the DES sources. The npAu layer is comprised of a 10 nm Ti adhesion layer, a 100 nm Au layer, and a (1.6 ± 0.1) µm Au and Ag alloy. Each film was deposited using a magnetron-based sputtering system (Denton Vacuum Explorer, Moorestown, NJ) with Au and Ag targets in DC and RF modes at 1.2 A and 300 W, respectively. The film was then de-alloyed in 65 % $HNO_3$ overnight to reveal the porous network seen in Figure 4. Volumetric solute transfer was confirmed by depositing an aqueous nitric acid solution containing 5 µg $g^{-1}$ (5 ppm) $CsNO_3$, followed by secondary ion mass spectrometry (SIMS) depth profiling to track the presence of $Cs^+$ ions as a function of film depth. The Cs was a non-radioactive surrogate for Am and is a great tracer for studying diffusion given the high ionization potential without changing the fluid properties or chemical composition.

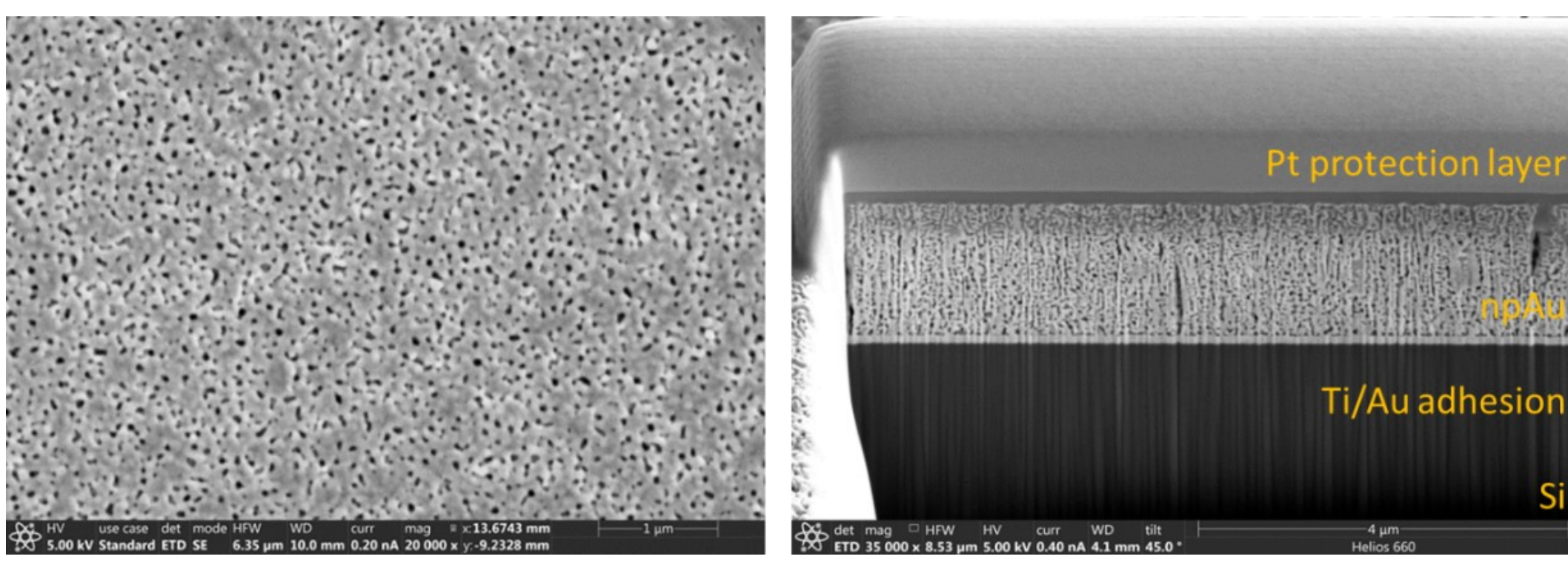

Figure 4 Scanning electron micrographs showing the surface structure of the nanoporous Au (npAu) film (left), and the cross section of the film after focused ion beam (FIB) cross sectioning (right). This particular npAu film was deposited onto Si wafers to facilitate mounting and analysis using advanced surface analytical tools. The average diameter of the pores on the surface of the films were measured to be (44 ± 13) nm with a porosity of 21 %, while those from the cross section were (47 ± 26) nm and 18 %, respectively, and shows the uniformity of the porous structure.

SIMS depth profiles were obtained using an IONTOF IV instrument (Münster, Germany) equipped with a 30 keV $Bi^+$ liquid metal ion source for both analysis and sputtering. Depth profiling was performed in non-interlaced mode, where 50 scans of analysis in pulsed beam mode and 75 scans of sputtering in continuous beam mode were operated separately per cycle (Figure *5*). To prevent pillar-like topography from changing the sputter rate, the sample was rotated by 90° after every cycle [16]. Crater depth measurements were performed using the Dektak XT stylus profilometer (Bruker Corporation, Tucson, AZ) equipped with a 2.5 µm radius stylus tip. As can be seen in Figure *5*, the distribution of the Cs was fairly uniform throughout the entire thickness of the film. $Cs^+$ and $Ag^+$ present past the Ti interface is due to a mixing effect where irradiation of the energetic beam leads to "knocking in" of elements deeper into the matrix. However, an order of magnitude drop in $Cs^+$ intensity signifies a clear change in layer composition. The main finding here is that the solution containing the tracer was able wick into the porous film. The npAu films were subsequently cut into 2.4 mm squares using a laser cutter and cleaned again using nitric acid to eliminate ambient carbon contamination before inkjet deposition.

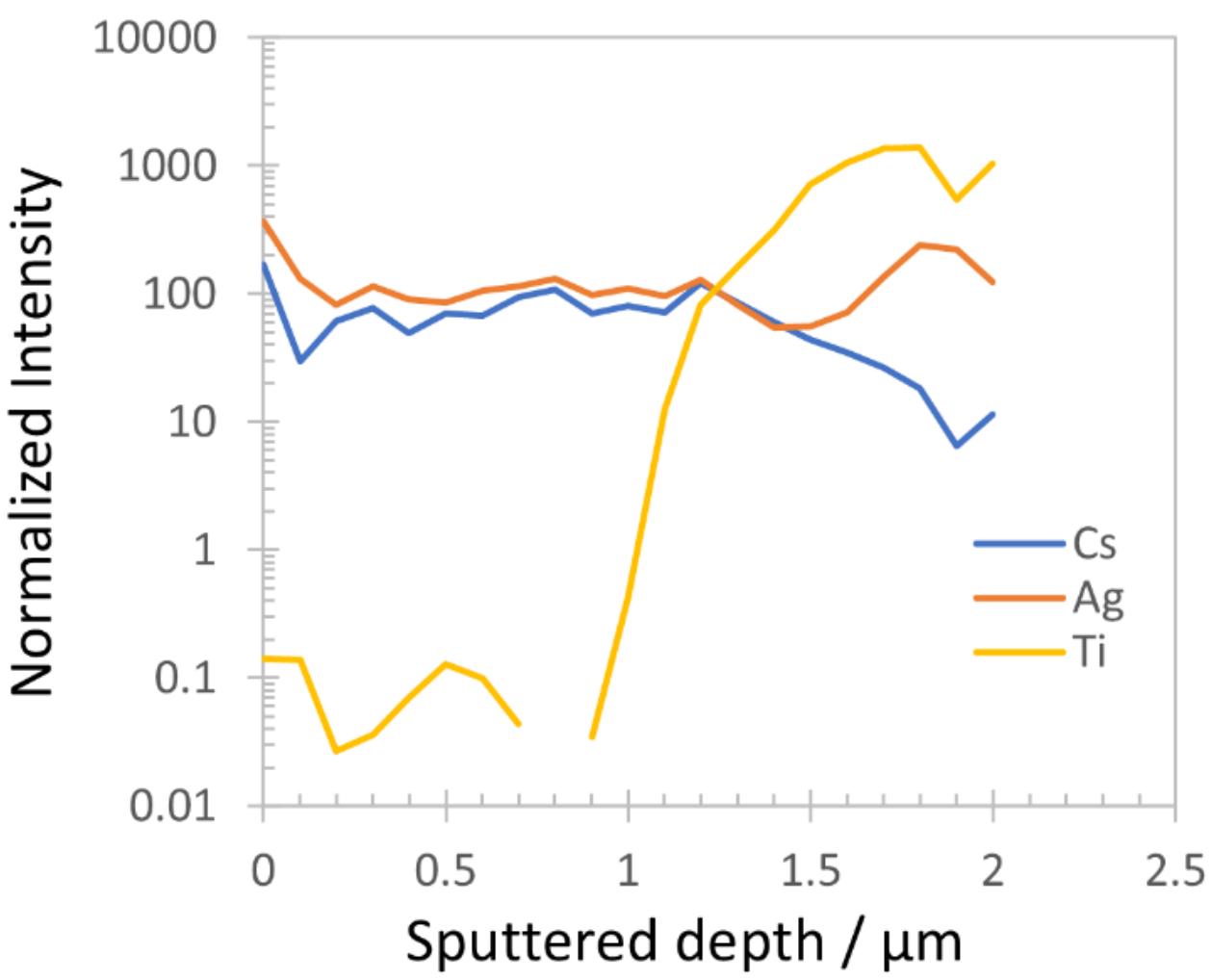

Figure 5 Secondary ion mass spectrometry (SIMS) depth profile showing the intensity of $Cs^+$, $Ag^+$, and $Ti^+$ ions inside the npAu film as a function of sputtered depth. The intensities were normalized to the $Au^+$ ions to correct for the preferential sputtering effect at the initial transient region. Note that the intensity is plotted on a log scale, and the position of the Ti interface occurs around a depth of 1.5 µm at 50 % of the maximum intensity.

### 2.3 Gravimetric Inkjet Dispensing

Inkjet dispensing of the Am-241 solution was performed using a metrology-level Jetlab 4 tabletop printer (Microfab Technologies, Plano, TX)* that incorporates an ultramicrobalance to calibrate drop mass before and after dispensing onto the npAu DES substrates. Previous work at

* Certain commercial equipment, instruments, or materials are identified in this paper to foster understanding. Such identification does not imply recommendation by the National Institute of Standards and Technology, nor does it imply that the materials or equipment identified are necessarily the best available for the purpose.

NIST has established procedures for measuring inkjet-dispensed masses with < 1 % relative combined standard uncertainty [17-19]. Both DES sources (0.01 mg to 0.1 mg) and liquid-scintillation (LS) sources (1 mg to 10 mg) were prepared by inkjet dispensing.

For the present work, additional steps were taken to mitigate evaporation effects on the measured dispensed mass. During mass calibration, the inkjet tip was lowered to a few mm above the surface of 1 mol/L $HNO_3$ contained in a gold-plated glass vessel with a ≈ 2 mm diameter opening. Previous tests, performed at multiple heights above the liquid to estimate residual evaporation, allowed correction for evaporation-in-flight of droplets travelling to the balance vessel.

To check for evaporative changes in the concentration of the Am-241 solution in the inkjet reservoir, LS sources were prepared from the solution before and after preparation of the DES sources by both the traditional pycnometer method (larger masses) and by inkjet deposition [20]. These LS sources served as a radionuclidic confirmation of inkjet gravimetry.

To test for potential losses of radioactive material due to stray satellite droplets of solution or leakage from the foil, the "placemat" method described previously [13] was employed. Each of six gold foils (2.4 mm square) was placed on a square of hydrogel during dispensing of larger (≈ 1 mg, 16 000 drops, 160 Bq) aliquots of the Am-241 solution; the foils for DES were also prepared over gel placemats, but the lower activities made detection of any stray material unlikely, so this dedicated study with higher activity was undertaken. After the foil sources dried, each foil and gel were placed in their own LS vials and counted in two commercial liquid scintillation counters to test for any activity lost from the foil during dispensing or drying.

Following drying of the inkjet-deposited solution on the nanoporous surface of the foil, each foil was folded in half and rolled flat using a manual roller press to encapsulate the source material [21]. Each foil was then coated on one side with a thin layer of In-Ga eutectic for thermal contact with the sensor and placed on the 2 mm by 1 mm gold-coated sample pad on a 5 mm by 5 mm TES chip.

### 2.4 Decay Energy Spectrometry

DES was performed using transition-edge sensors (TES) similar to those used for plutonium isotopic ratio studies [22]. Since the goal of this work is absolute massic activity of radionuclides that are generally well-separated in decay energy, the focus was on simple, quantitative source preparation, with modest (for cryogenic sensors) energy resolution of about 5 keV. A dilution refrigerator with temperature control was used to maintain a base temperature of 50 mK (27 μK root mean square stability). A total of 5 foil-encapsulated Am-241 sources and 2 blank foils, mounted on the sample pads of TES chips, were measured on multiple occasions. Each chip was voltage-biased to approximately 20 % of the TES normal resistance. Data were acquired with commercial SQUID amplifiers and analog-to-digital converters (ADCs). Continuous signals were recorded at a sample rate of $10^5$ $s^{-1}$.

Typical pulses had exponential time constants of approximately 0.2 ms on the rising edge and fall times of 25 ms on the falling edge. Pulse processing consisted of a live-timed [23-25] counting algorithm that required an absence of triggers within a fixed time before a pulse (traditional live-timing using extending, or paralyzable, dead time [26]) and after a pulse (to ensure clean pulses for spectrometry) for the time to be considered live. Dead-times were chosen to be long enough to reduce peak-broadening due to pileup while short enough not to significantly increase counting uncertainty. Since we fit the tail of the previous pulse as baseline under the pulse-under-analysis, we did not need too long a dead-time. Typically, live time fraction was about 80 %.

The offline pulse processing implementation followed a series of steps. Refer to Figure 3. First, the data stream was convolved with a "trigger filter", in the form of a step function of adjustable length (typically 10 samples) chosen to optimize low energy sensitivity and timing jitter. A threshold is set on the running convolution to set a trigger. A holdoff of 0.5 ms was applied after a trigger to avoid double triggering. Second, pulses with sufficient time after the previous trigger (imposed deadtime A) and time until the next trigger (imposed deadtime B) are accepted for energy analysis (though all pulses generate dead-times). If a pulse is found within deadtime B after the pulse under analysis, deadtime B is extended until the next pulse in the record. Typical values for imposed deadtimes were 100 ms to 400 ms for deadtime A and 10 ms for deadtime B. Third, an "energy filter" is applied to each pulse to extract a pulse amplitude. The energy filter is an average pulse amplitude starting at a fixed time after each trigger. Typically, this was set at 5 ms after the trigger and averaged over 30 samples. This method gave similar energy resolution to an optimal filter but was less susceptible to pulse shape variability, particularly in the rising edge. Finally, a linear energy calibration is implemented, based on the clear Am-241 Q peak. The linearity of the calibration was checked by measuring the spectrum from a Co-57 source mounted outside the cryostat. A subset (about 0.3 %) of pulses was identified where a gamma ray escaping from the npAu foil interacts with the rest of the TES chip island directly; these pulses have a faster-rising edge and a larger proportion of the total pulse area in the first 1 ms of the data. The spectroscopic information from these pulses is reduced. Such events show up in the spectrum as a low-energy tail on the Q-value peak; since the only $\gamma$-ray emitting radionuclide identified in this sample was Am-241, those events can be assigned as Am-241 decays. In measurements with multiple nuclides, these pulses will not be so easily accounted for, so we intend to alter the TES design to reduce the rate of these interactions by at least 10 times.

The activity in a given region of interest (ROI) is the sum of counts in that ROI divided by the live time of the histogram.

To check that there were no hidden dead-times or double triggering, a histogram of arrival time differences between subsequent triggers (Figure 6) was created (see Reference [27] for motivation). The data follow the expected exponential distribution for random radioactive decay events.

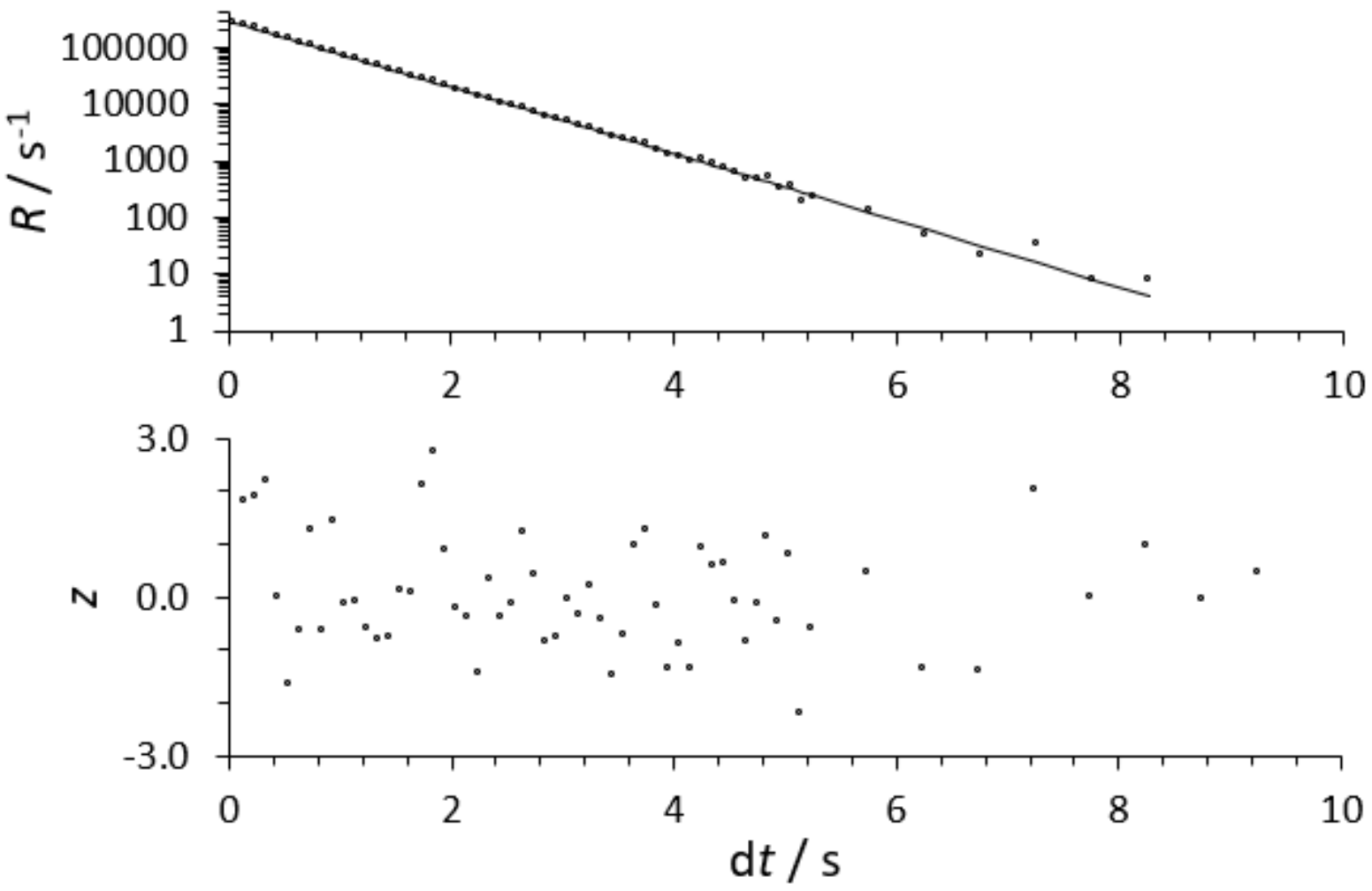

Figure 6 Top graph is time-interval distribution of triggered pulses, before application of live-timing, where *R* is the count rate (counts per bin width). Solid line is an exponential fit to the data. Bottom graph is z score (residual / counting uncertainty). The first point, at $t$ = 5 ms is not shown, having a value of $z$ = -8.8, due to trigger length. The live-timed counting method used extending deadtimes much longer than 5 ms (typically 200 ms) to ensure accurate absolute counting. The standard deviation of the $z$ scores shown is 1.07.

The resulting DES spectrum was compared to Geant4 [28] Monte Carlo simulations to search for any radioactive impurities within the Am-241 energy region of interest (ROI) shown in Figure 7. As no impurities were found, the Am-241 count rate was taken as the sum of live counts in the Am-241 ROI divided by the live time. Had impurities been present, the impurity spectrum could be fit using the simulated spectra to subtract the Am-241 contribution to the impurity ROI. The best energy resolution achieved was characterized using a convolution of a Gaussian with 3.3 keV standard deviation and a left-sided exponential with a decay constant of 4 keV. The 3.3 keV is consistent with measurements of noise records (sections of the TES data stream having no pulses). Beyond the energy range of Figure 7, a pileup peak was observed at around 11 MeV with a count rate relative to that of the Am-241 region of interest of $7.4 \cdot 10^{-4}$, consistent with expectations for an Am-241 decay rate of 1.35 Bq and a trigger holdoff of 0.5 ms.

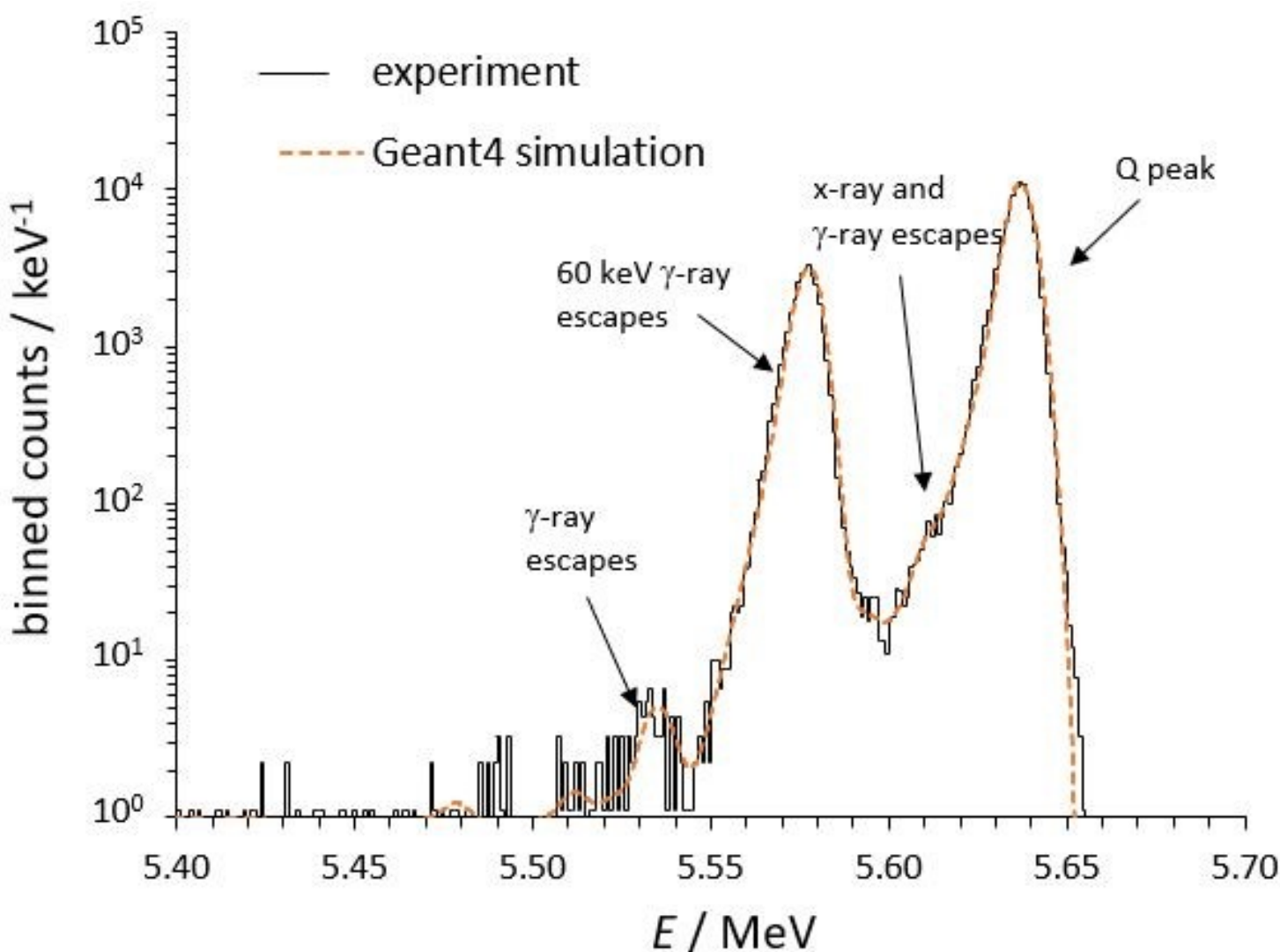

Figure 7 Experimental Am-241 spectrum (25 μm thick foil; 45 h measurement) with fit using Geant4-simulated spectra (Gaussian width and left exponential tail as free parameters, applied to the whole spectrum). The Am-241 decay energy, Q, and escape peaks from $\gamma$ rays and x rays are indicated.

## 2.5 Validation measurements

The solution was assayed for massic activity of Am-241 before and after the DES experiment using well-established non-spectroscopic primary methods based on 4πα liquid scintillation (LS) counting. LS counting was carried out on two commercial instruments, one operated with an open energy window for absolute counting and one operated in a low-background mode, and on a custom-built three photomultiplier tube counter (often referred to as a "TDCR system", even when the triple-to-double coincidence ratio method is not employed) [29]. γ-ray spectrometry [30] measurements were performed to check for photon-emitting radionuclide impurities.

The low-background LS instrument, featuring alpha-beta discrimination, was used primarily for the placemat measurements, due to its high sensitivity. The open-window LS counter was used to confirm the Am-241 massic activity before and after the inkjet dispensing of DES sources to validate that the massic activity did not change (e.g., due to evaporation in the inkjet reservoir). A subset of sources was counted on the TDCR counter to confirm the open-window counter results. 4π LS counting is a well-established method for non-spectroscopic primary standardization of activity when the nuclide and purity of a sample is known and was used for live-timed absolute counting [29, 31].

3. Results

3.1 Background and impurities

Two blank foils were measured by DES for 40 000 s and 160 000 s, with 1 count and 0 counts above 0.5 MeV, respectively, corresponding to negligible background for Am-241 (Figure 8). The trigger threshold was set based on signal-to-noise for each spectrum, corresponding to about 15 keV. No evidence of Pu-241 decay (beta endpoint 20.5 keV) was seen. The beta background rate was about $6 \cdot 10^{-4}$ s$^{-1}$, which is about $10^4$ times lower than that for LS. Gamma-ray spectrometry measurements were carried out for 70 h and no impurities were found. The most likely impurities (Pu isotopes) produced few to no γ-rays, so the DES spectra provide a stronger impurity constraint. An energy calibration for the blank foil was achieved using the 122 keV $\gamma$-ray line from a $^{57}$Co source placed external to the cryostat.

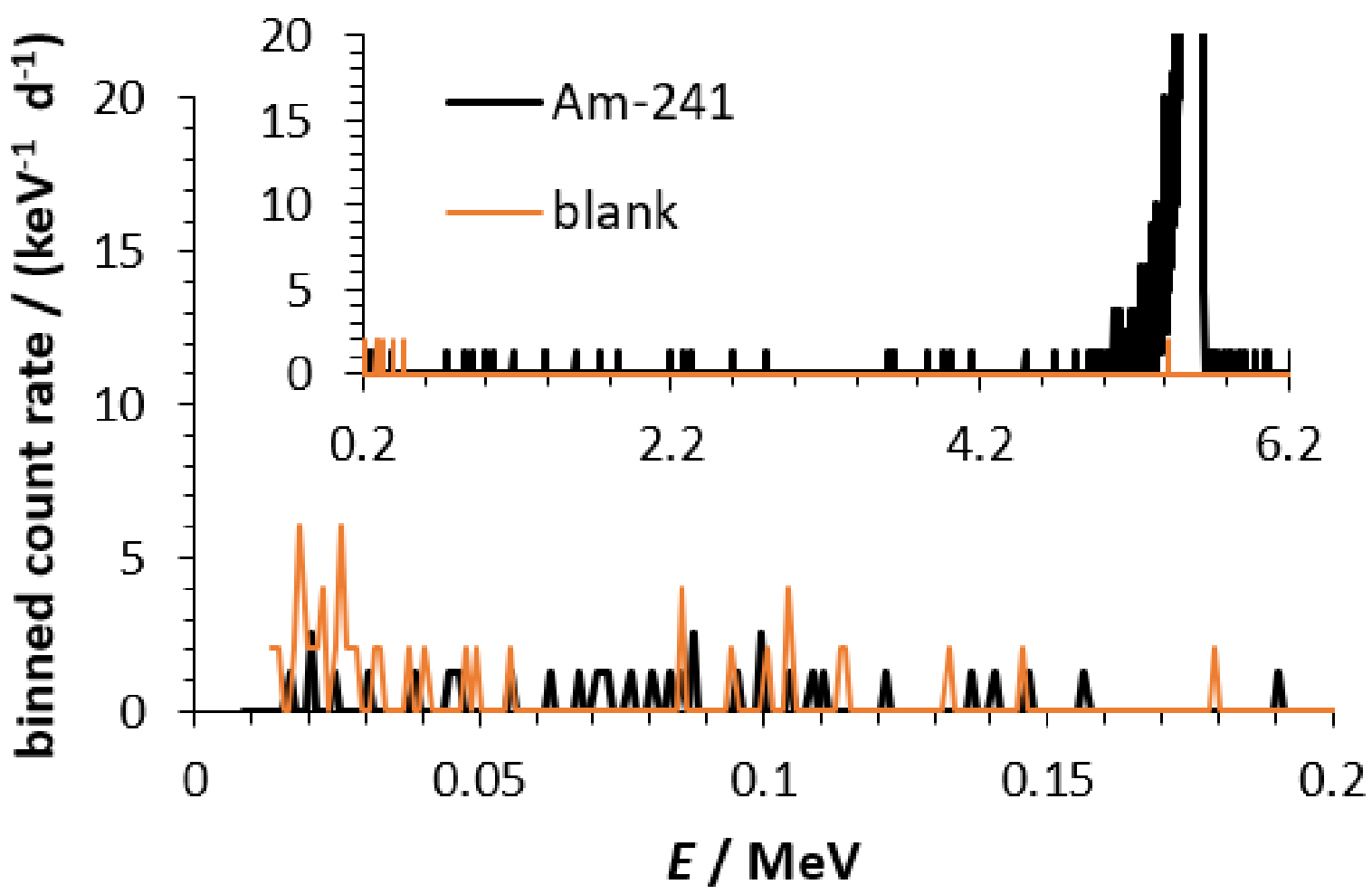

Figure 8 DES spectra for an Am-241 source and for a blank foil measured concurrently on separate TES chips. The noise cutoffs are about 10 keV and 15 keV for the Am-241 and blank, respectively. No evidence of Pu-241 (20.5 keV endpoint) is seen. The count rate between 20 keV and 200 keV was $(4.7 \pm 0.8) \cdot 10^{-4}$ s$^{-1}$ and (6.8 ± 1.3)$\cdot$ $10^{-4}$ s$^{-1}$ for the Am-241 and blank, respectively, indicating no beta-particle emitting impurities in that region. In the Am-241 region, there was 1 count in 43000 s of measurement time for the blank, with a rate $(2.4 \pm 0.2) \cdot 10^{-5}$ s$^{-1}$. A second blank was run for longer with no counts, giving an upper limit of $\approx 6 \cdot 10^{-6}$ s$^{-1}$.

The resolving power for DES was about 150 times that for LS, allowing for detection of potential radionuclide impurities by DES (Figure 9). Figure 10 shows simulated spectra containing Pu-240 ($Q$-value = 5.256 MeV) and Pu-238 (5.593 keV) impurities at activity ratios relative to Am-241 of 0.02 % and 0.15 %, respectively. The spectral evidence of these low-level impurities illustrates the value of this spectroscopic method for, e.g., characterizing Standard Reference Materials (SRM), which typically have > 0.15 % uncertainties in their combined standard uncertainty for the activity of the main radionuclide. No evidence of low-energy beta decays from Pu-241 was observed, though the trigger threshold is near the endpoint energy.

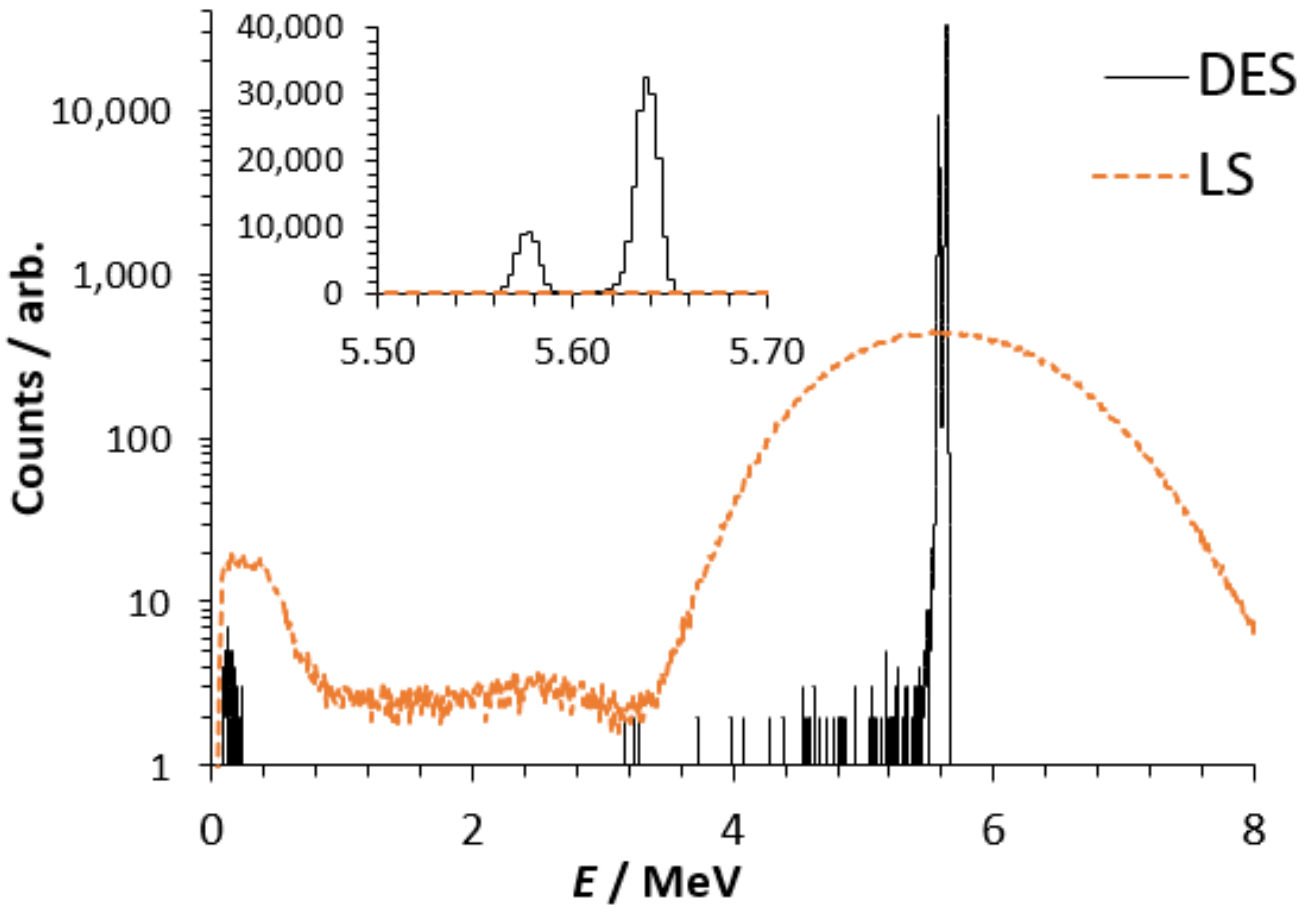

Figure 9 Comparison of DES and LS spectra for Am-241. (Note log scale vertical scale, linear on inset). Both detectors have ≈ 100 % counting efficiency, but DES has much higher energy resolution. Note, the LS detector response is highly non-linear and differs for alphas, betas, and gammas. Here the bin energies have been scaled linearly to align the alpha peak to 5.6 MeV. The two spectra are scaled to contain the same integral number of counts in the spectrum. No background subtraction has been made, such that low-energy background is present (also seen in blank samples).

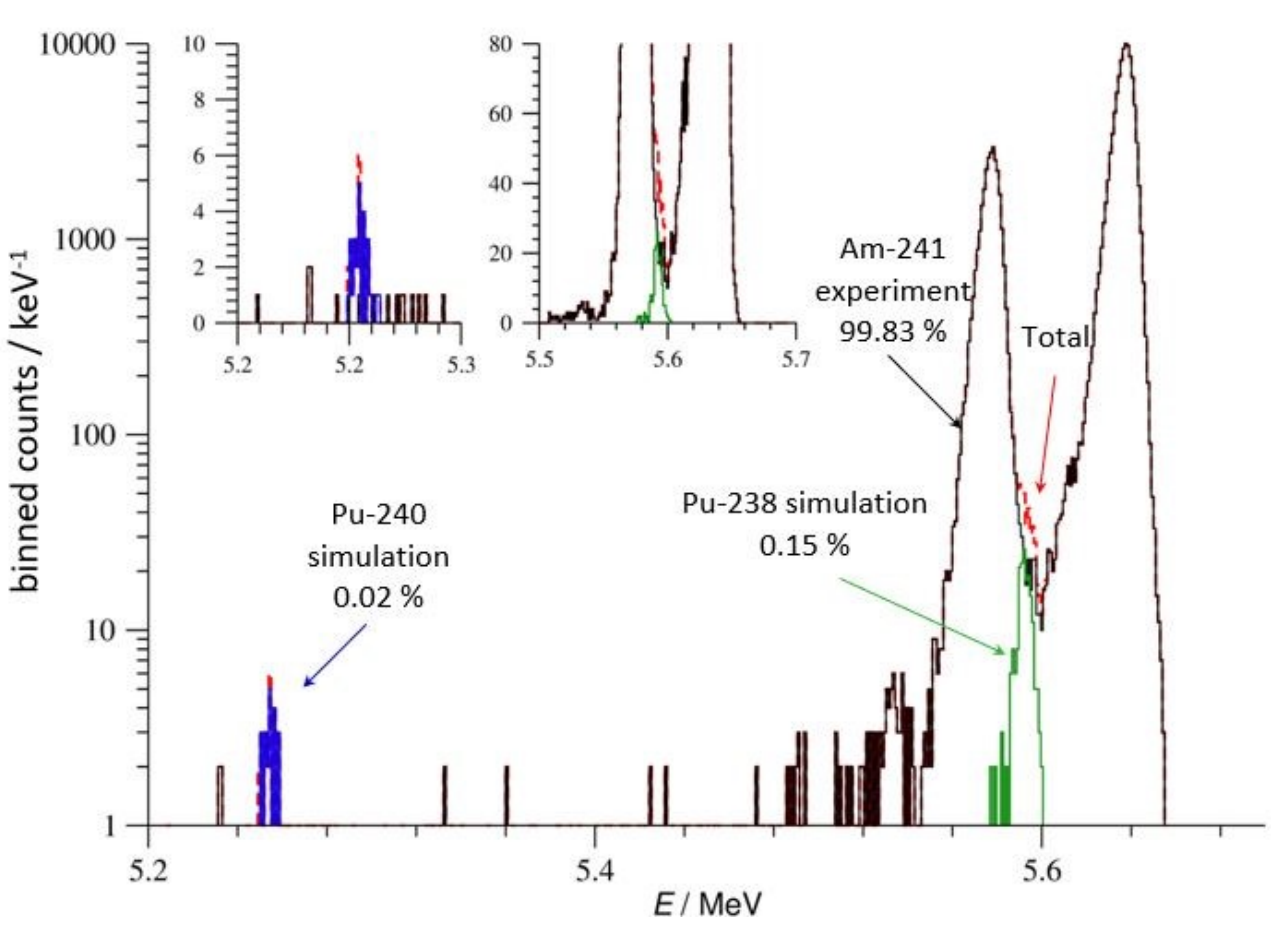

Figure 10 Impurity search: Same experimental Am-241 spectrum from Figure *7* (black ≈126 000 counts) with the addition of Geant4 simulated impurities of Pu-240 (blue, 25 counts, or 0.02 % of total) and Pu-238 (green, 189 counts, or 0.15 % of total) to illustrate detectability in the composite spectrum (red). No evidence of radionuclide impurities was found in the experimental spectrum, with detection limits at a 95 % coverage level for activity relative to Am-241 of 0.005 % and 0.09 % for Pu-240 and Pu-238, respectively.

## 3.2 Placemat study

The average and standard deviation of the distribution of the relative activity on the placemats vs. foils for the 6 sources was (0.017 $\pm$ 0.018) % (Figure 11). The highest value was 0.053 %, which is a 5 times improvement over the earlier hand-dispensed microdrop method [13]. Based on these results, a conservative uncertainty of 0.05 % was assumed due to material loss during the quantitative source preparation for the DES measurements.

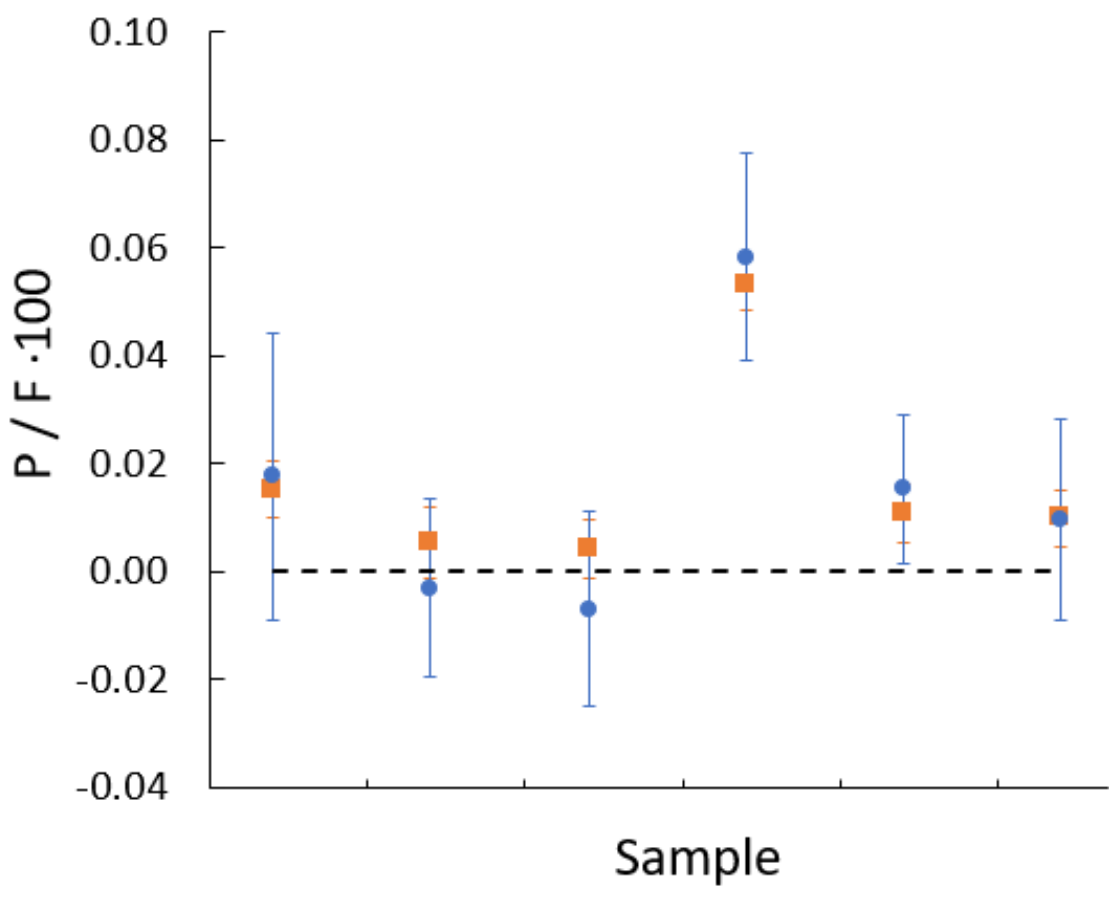

Figure 11 Results of placemat test showing the ratio of (background-subtracted) counts on the placemat (P) to that on the foil (F), as measured in two different LS counters. For the low background LS counter (orange squares), the statistical uncertainty on the net count rate (*P*) for the placemats was about 0.004 $s^{-1}$, or 0.002 % of the foil count rate (*F*).

### 3.3 Massic activity

The weighted average activity for 4 DES sources was (1.357 ± 0.007) Bq, which combined with the dispensed mass from the inkjet, gives a massic activity of 145.7 kBq $g^{-1}$ with relative combined standard uncertainty $u_c$ = 0.55 % (Table 1). The first DES source in the series gave a positively biased activity that is explained by a concentrating first aliquot effect. We have investigated this effect further in a series of follow-up experiments, but in the present work, to avoid uncertainties introduced by any correction, the first sample was simply excluded from the massic activity analysis. The major sources of uncertainty were counting statistics (0.49 %) and the uncertainty in the dispensed mass (0.21 %), the latter due mostly to regression of mass vs. time (including reproducibility among weighing), standard deviation among 4 analysis algorithms, balance linearity, and evaporation-in-flight correction. The relative counting statistics uncertainty (0.49 %) was taken as the uncertainty in the weighted mean count rate for 4 Am-241 samples (same number of drops of active solution), where the uncertainty for a measurement on one source was the Poisson counting uncertainty within the ROI. For comparison, the "external uncertainty," calculated as the weighted standard deviation for the 4 sources was 0.36 %, indicating that the sources were statistically identical. The source mass uncertainty was primarily due to the calibration of the mass per drop from the inkjet [17]. Despite the application of forward and backward extending dead-time, a small pileup peak was observed at around 11.2 MeV, containing about 0.06 % of all events and assigned as single Am-241 decays, with an associated uncertainty of 0.06 %.  As an estimate on the uncertainty due to unobserved impurities, a standard uncertainty was estimated to be 0.05 %, based on the detection limit of 0.09 % for Pu-238.

Table 1. Uncertainty analysis for DES massic activity measurement where $u_i$ are the relative uncertainty in the final massic activity of the solution due to each of the input quantities. The $u_i$ values are added in quadrature to determine $u_c$ [32].

| **Component** | **$u_i$ / %** |
| --- | --- |
| Counting statistics | 0.49 |
| Live time | 0.10 |
| Background | 0.00 |
| Pulse assignment | 0.06 |
| Mass | 0.21 |
| Limit on material loss | 0.05 |
| Limit on impurities | 0.05 |
| **$u_c$ / %** | **0.55** |

The standard LS value taken for the massic activity was that of the TDCR system for 5 LS sources prepared by pycnometer, with an average massic activity of 144.85 kBq g$^{-1}$ and $u_c$ = 0.33 %. All of the LS and DES measurements resulted in consistent massic activities for the solution (Figure 12). No change in massic activity was observed during the experiment from evaporation or other effects.

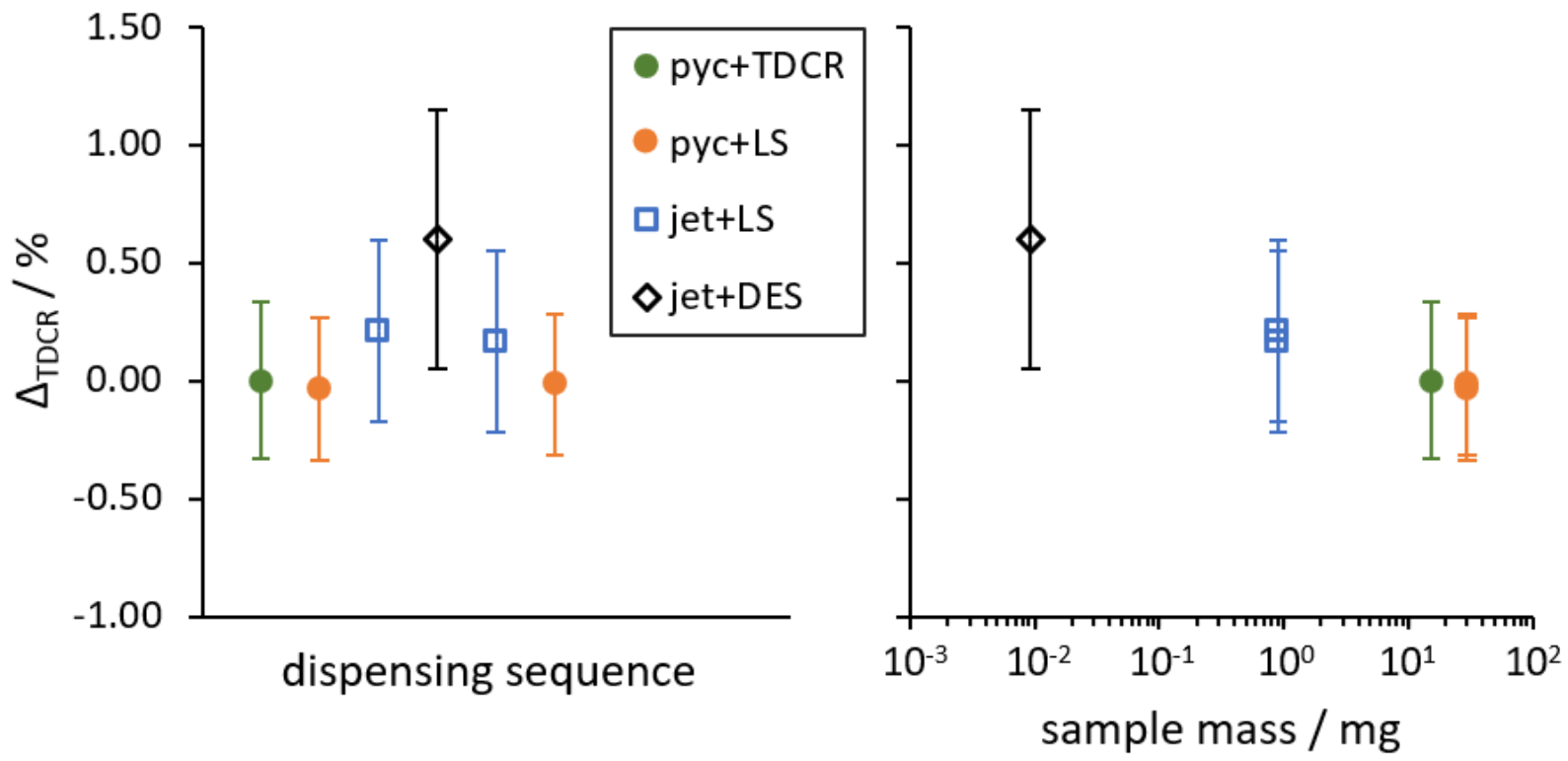

Figure 12 Relative difference $\Delta_{TDCR}$ of massic activity from the average of the standard method (pycnometer + TDCR). Bars are combined standard uncertainties on each measurement method.

**Left:** results plotted in time order of source dispensing; no change in massic activity from evaporation or other effects was observed over time. **Right:** results plotted against aliquot mass per sample to highlight the good agreement even for very small sample masses achievable by DES.

## 4. Discussion

In this work we successfully quantify Am-241 massic activity while spectrally excluding the presence of the nearby Pu-240 and Pu-238 isotopes at the 0.02 % to 0.15 % level. Geant4-based radiation transport simulation results validate understanding of the DES and predict sensitivity to these radioactive impurities. These impurities are invisible to γ-ray spectrometry because they produce only weak photon emission above 20 keV and would require chemical processing to detect by traditional alpha spectrometry. Further, since Pu-241 and Am-241 are isobars, they can interfere in mass spectrometry but should be resolvable by DES (decay energies of 21 keV and 5637 keV, respectively), with somewhat improved noise threshold. This capability enables primary measurement of activity for multiple radionuclides in one assay on drop-dried sources, without the need for chemical processing and the attendant chemical tracers.

Liberation from chemical separations and efficiency tracers brings benefits beyond faster throughput and lower cost. In traditional methods, major sources of uncertainty in the massic activity measurement include chemical separation and detection efficiency, along with ambiguities from spectral interferences from isotopic impurities. For example, recent measurements of a Pa-231 solution had a goal of $< 0.2$ % uncertainty, but the results had uncertainty $> 3$ % due to these issues [33]. If DES were applied to this case, sources would be made and counted with 100 % efficiency and spectral resolution would cleanly separate Pa-231 from an Ac-227 impurity, resulting in predicted uncertainties within the goal.

## 5. Conclusion

An advantage of the DES method is that it can be used for both massic activity and relative activity. The measurements reported here are quantitative in activity and spectroscopic, enabling a new capability of multi-nuclide absolute activity. Through inkjet gravimetry, DES sources have been prepared with precise and radiometrically validated links to a stock solution. This paper thus presents a DES-determined massic activity, validated with another primary method and achieving relative combined standard uncertainties that are competitive with mature techniques in radionuclide metrology.

Longer measurements and higher count rates can be employed to reduce statistical uncertainties. With a TES optimized for higher heat flow, to shorten the pulse recovery time, the activity per source could be increased, resulting in lower counting uncertainties. In future quantitative DES work, the energy resolution, which is important to the sensitivity to both alpha and beta impurities, may be improved by 4 times or more based on the highest resolution DES results in the literature [10] by reducing electrical and vibrational noise and decreasing potential crystal size in dried deposits. A combination of quantitative sources and highest-resolution sources could be measured in a single experiment.

For environmental samples, increased sample size could be achieved by cycling inkjet and drying steps. For decay chains, short-lived chain members with half-lives on the order of the TES pulse recover time ($\approx$10 ms) or shorter can be accounted for by the traditional method of calculating pulse survival probabilities, , and potentially using spectroscopic quantification of piled-up pulses.

Efforts are underway to decrease the uncertainty in sample mass using a new balance being developed at NIST. This device incorporates the dispenser into the balance itself and realizes mass traceable directly to electrical measurements. For 1 mg masses of gravimetrically diluted solution, expected uncertainties are of order 0.1 % [37].

## 6. Acknowledgements

The authors acknowledge Mark Croce and Katherine Schreiber of Los Alamos National Laboratory and Geon-Bo Kim of Laurence Livermore National Laboratory for sharing source preparation methods. RF acknowledges the NIST Radiation Physics Building modernization team for their support. This work has been funded by the Innovations in Measurement Science program at NIST.